\documentclass[12pt]{iopart}
\usepackage{graphicx}
\usepackage{iopams}
\newcommand{\be}{\begin{equation}}
\newcommand{\ee}{\end{equation}}

\begin{document}
\title[Critical behavior of the geometrical spin clusters and interfaces in the 2D TBIM]{Critical behavior of the geometrical spin clusters and
interfaces in the two-dimensional thermalized bond Ising model}

\author{S.\ Davatolhagh,$^1$ M.\ Moshfeghian$^1$ and A.\ A.\ Saberi$^{2,3}$}
\address{$^1$Department of Physics, College of Sciences, Shiraz
 University, Shiraz 71454, Iran}
\address{$^2$Institut f\"{u}r Theoretische Physik,
Universit\"{a}t Zu K\"{o}ln, Z\"{u}lpiche Str. 77, 50937 K\"{o}ln,
Germany}
\address{$^3$School of Physics, Institute for Research in Fundamental Sciences (IPM), PO Box 19395-5531, Tehran, Iran}
\ead{davatolhagh@susc.ac.ir}

\date{\today}

\begin{abstract}
The fractal dimensions and the percolation exponents of the
geometrical spin clusters of like sign at criticality, are obtained
numerically for an Ising model with temperature-dependent annealed
bond dilution, also known as the thermalized bond Ising model
(TBIM), in two dimensions. For this purpose, a modified Wolff
single-cluster Monte Carlo simulation is used to generate
equilibrium spin configurations on square lattices in the critical
region. A tie-breaking rule is employed to identify non-intersecting
spin cluster boundaries along the edges of the dual lattice. The
values obtained for the fractal dimensions of the spanning
geometrical clusters $D_{c}$, and their interfaces $D_{I}$, are in
perfect agreement with those reported for the standard
two-dimensional ferromagnetic Ising model. Furthermore, the variance
of the winding angles, results in a diffusivity $\kappa=3$ for the
two-dimensional thermalized bond Ising model, thus placing it in the
universality class of the regular Ising model. A finite-size scaling
analysis of the largest geometrical clusters, results in a reliable
estimation of the critical percolation exponents for the geometrical
clusters in the limit of an infinite lattice size. The percolation
exponents thus obtained, are also found to be consistent with those
reported for the regular Ising model. These consistencies are
explained in terms of the Fisher renormalization relations, which
express the thermodynamic critical exponents of systems with
annealed bond dilution in terms of those of the regular model
system.\end{abstract}
 \maketitle
\noindent {\bf keywords:} Stochastic Loewner evolution, Critical exponents and amplitudes (theory), Classical Monte Carlo simulations

\section{Introduction}
It is well known that the geometrical spin clusters (i.e., the
clusters composed of the neighboring spins of the same sign), undergo a
percolation transition at the thermodynamic critical temperature
$T_c$ in the two-dimensional Ising model \cite{CK80}. This type of
behavior is believed to be generally valid for a variety of
two-dimensional critical models that undergo a continuous phase
transition such as the $q$-state Potts model ($q\leq 4$) \cite{For},
which is a $q$-state generalization of the Ising model ($q=2$)
\cite{Pot}. The formation of spanning spin clusters and their
characteristic percolation exponents, can also be used to
characterize the universality class of the corresponding thermal
phase transition. Furthermore, at the critical point, the spanning
cluster is a scale-invariant fractal object whose fractal dimensions
uniquely specify the universality class of the associated continuous
phase transition \cite{GC}.

The critical behavior of a great number of statistical models in two
spatial dimensions, has been investigated by the conformal field
theory \cite{AMB}. The conformal invariance property, refers to
the invariance under coordinate transformations through which the angles
between the crossing lines in the $z$-plane do not change. From this
point of view, the cluster boundaries in the two-dimensional
critical systems are considered as conformally invariant curves, and
different characteristics such as their fractal dimensions are
obtained. Indeed the fractal geometry is a useful mathematical tool
for the characterization of a great many complex configurations.
Self-similarity is the most important characteristic of such fractal
objects. It is well known that the most widely studied statistical
models in the condensed matter physics such as the Ising model, its
$q$-state generalization the Potts model at criticality as well as
the many critical geometrical phenomena exhibited by the various
percolation models, consist of fractal lines
\cite{GC,Ca1,Co,Du,NC,N}.

More recently, the spin cluster boundaries (interfaces) in the
two-dimensional critical models, have been investigated rigorously
using the method of Stochastic Loewner Evolution (SLE), in which the
motion of a random walker along the cluster boundary in the
upper-half complex plane in the continuum limit is specified by the
Loewner dynamics \be \frac{\partial g_t(z)}{\partial
t}=\frac{2}{g_t(z)-\zeta_t}\,,\label{eq1.0}\ee which contains a
Brownian term $\zeta_t=\sqrt{\kappa}B_t$ whose amplitude is given by
the SLE parameter $\kappa$, also known as the diffusivity
\cite{Sch}. The function $g_t$ maps a parametric curve $\gamma_t$ in 
the upper-half complex plane onto the real axis. Thus, given a real 
function $\zeta_t$ and using the initial condition $g_0 = z$, the 
Loewner differential equation (\ref{eq1.0}) determines the corresponding 
trace $\gamma_t$ in the upper-half $z$-plane.  The larger the diffusivity, $\kappa$, the more is the
deviation from a straight line. Indeed the nature of the SLE traces
change with the diffusivity: for the range of values
$0\leq\kappa<4$---a range that includes $\kappa =3$ characterizing
the boundaries of the geometrical clusters (of like sign) in the
critical two-dimensional Ising model---the SLE traces are
nonintersecting simple curves, while for $4\leq\kappa\leq8$ the
curves possess double points with possible self-touching (but no
crossing), and for $\kappa
> 8$ they become space-filling \cite{Sch,BB,Ca2}. Thus, the critical
fractal dimension of the interfaces $D_{I}$ introduced by this
theory are model dependent. The relation between $D_{I}$ and
$\kappa$ is given by $D_{I}= 1 + \frac{\kappa}{8}$  \cite{BV}. The
exact values are $\kappa = 3$ and $D_{I}=\frac{11}{8}$ for the geometric
spin cluster boundaries of the two-dimensional regular Ising model,
as obtained analytically through various methods \cite{SS2}. Like
the thermodynamic critical exponents in statistical mechanics,
$\kappa$ can divide different models into universality classes
\cite{Ca2}. The main difference between $\kappa$ and the thermal
critical exponents, is due to the method of definition---one being
thermodynamic and the other geometrical in nature. Normally, the
characterization of the universality class, requires a minimum of
two thermodynamic critical exponents. We note that in the case of
two-dimensional critical systems, a single parameter, $\kappa$,
appears to be sufficient to specify the universality class. This may
be attributed to the fact that of the two specifications of the
model that significantly influence its critical behavior, i.e.\
space dimensions $d$ and the order parameter dimensions $n$, one is
held fixed at $d=2$. Hence, the SLE diffusivity $\kappa$ alone can
be used to specify the universality class of the two-dimensional
critical systems.

In the following we focus on the thermalized bond Ising model (TBIM)
in two dimensions and investigate the behavior of its geometrical
spin clusters (i.e., spin clusters of like sign) and their external
boundaries (interfaces) at criticality. A  Wolff single-cluster
Monte Carlo algorithm is used to generate configurations at and near
the criticality on square lattices, and a tie-breaking rule is used
to identify non-intersecting geometrical cluster boundaries along
the edges of the dual lattice. The rest of this paper is organized
as follows. In section 2, the thermalized-bond Ising model is
briefly reviewed, focussing on its thermodynamic critical behavior.
The method of simulation and the finite size scaling
procedures---employed to extrapolate the results obtained for finite
lattices to the thermodynamic limit---are explained in section 3. The
results are presented and discussed in section 4, and the paper is
concluded with a summary in section 5.

\section{The model system}
\begin{figure}
\includegraphics{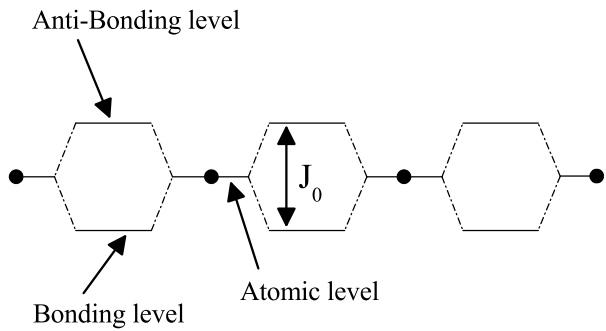}
\caption{Schematic illustration of the electronic energy states of
the covalent
 bonds linking a chain of atoms. The energy gap between the bonding and the
 antibonding level is denoted by $J_0$.} \label{figure1}
\end{figure}

The thermalized bond Ising model (TBIM), is a bond-diluted Ising
model with a temperature dependent bond concentration, in which
every covalent bond linking a nearest neighbor pair of atoms is
allowed thermally induced electronic transitions between bonding and
anti bonding electronic states \cite{DSB}. Hence, it can be regarded
as containing {\em annealed} bond defects with a temperature dependent
concentration. Each bond at every instant is characterized by a
coupling constant $J_{ij}= 0, J_0$, such that zero corresponds to a
broken bond (anti-bonding electronic state), while $J_0$ means an
attractive coupling between the two atoms (bonding electronic
state), as illustrated schematically in figure \ref{figure1}.
Denoting the thermally averaged bond concentration by $p_b$,
$(1-p_b)$ must therefore represent the concentration of the broken
bonds due to thermal excitations. To keep the analysis simple, the
covalent bonds are treated as independent two-level systems with
energy gap $J_0$, as sketched in figure \ref{figure1} \cite{DSB}.
The ratio of the bonds to the broken bonds in equilibrium, is given
by the ratio of the corresponding Boltzmann factors
$p_b/(1-p_b)=\exp(\beta J_0)$, or \be p_b = 1/(1+\rme^{-\beta
J_0})\label{eq1.1} \ee where $\beta=1/k_BT$ is the reciprocal
temperature and $k_B$ is the Boltzmann constant. The bond
distribution function for the thermalized-bond model system is of
the form \be P_{J_{ij}}(\beta) = p_b\;\delta_{J_{ij},J_0} +
(1-p_b)\;\delta_{J_{ij},0}\label{eq1.2}\ee where, $p_b$ is given by
equation (\ref{eq1.1}), and $\delta$ denotes the Kronecker delta
\cite{TB}. Hence, the Hamiltonian of the system can be formally
defined by \be H = \sum_{\langle i,j\rangle} J_{ij} S_i S_j \ee
where $S_i=\pm 1$ is an Ising spin, and the sum is over all
nearest-neighbor pairs.

It is well known that mapping the regular Ising model onto the
equivalent correlated percolation problem introduces the bond
probability \be P_r = 1-\rme^{-2\beta J}\label{eq2.1} \ee for the
critical droplets (the Fortuin-Kastelyn clusters) of the regular
Ising model \cite{CK80, FK}. The combination of equation
(\ref{eq1.1}) and equation (\ref{eq2.1}), together with the choice
$J_0=2J$, which is also the energy gap between parallel and
antiparallel spins in the regular Ising model, results in a compound
bond probability $P_{\rm TBIM}=P_r p_b$ for our thermalized-bond
model system:  \be P_{\rm TBIM} = (1-\rme^{-2\beta
J})/(1+\rme^{-2\beta J})=\tanh(\beta J).\label{eq2.2}\ee Equation
(\ref{eq2.2}) represents the bond probability used in our
single-cluster update MC simulations.

The thermodynamic critical behavior of the TBIM, has been studied before in two \cite{DM}, and three
dimensions \cite{DSB}. In two dimensions, the critical temperature is estimated to be
$T_{c}= 1.4897(3)$ \cite{DM}, which is
lower than the critical temperature of the regular Ising model. The
lowering of the transition point is expected in the light of the
annealed bond disorder present. However, the thermal critical exponents are found to be unchanged,
within statistical errors. As for the three-dimensional TBIM, the thermal critical exponents are found to change consistent with the Fisher renormalization relations, as the specific heat exponent of the regular Ising model in three dimensions, $\alpha_r\simeq 0.11$, is finite and positive.

\section{Simulation method and finite-size scaling}
\begin{figure}
\includegraphics{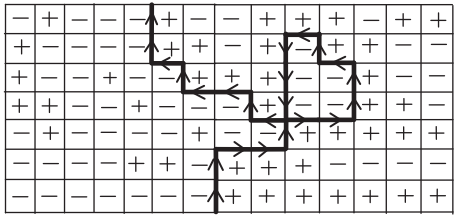}
\caption{An schematic illustration of defining the domain boundaries
in two-dimensional TBIM, on a square lattice. Figure shows the dual
of the original square lattice including a spin configuration, with
fixed boundary conditions for the bottom end, antiperiodic on sides,
and free for the top end. The non-intersecting interface (shown by
arrows) is generated using a turn right tie-breaking rule.}
\label{figure2}
\end{figure}

As pointed out in the introduction, the critical behavior of the
geometrical spin clusters and interfaces in the two-dimensional
TBIM, is the main goal of this paper. For consistency with the
postulates of SLE at $T_{c}$, we have considered the model on strips
of size $L_{x}\times L_{y}$, where the length of the strip $L_{x}$
is taken to be much larger than its width $L_{y}=L$ with an aspect
ratio $L_{x}/L_{y}=8$. The boundary conditions used for
simulations, are fixed for the lower boundary (real axis),
antiperiodic for the two sides, and free for the upper boundary of
the system, as shown schematically in figure \ref{figure2}. Using a
single-cluster update algorithm (Wolff's algorithm) \cite{UW} for
the two-dimensional TBIM on square lattice, we generated equilibrium
spin configurations at and near the critical point $T_c$. A typical run consisted of several weeks
of the CPU time on a single processor computer. Initially,
the system was allowed $2\times10^{3}\,L$ equilibration Monte Carlo
steps (MCS), and then the data points were accumulated by averaging
over $2\times 10^{2}\,L$ configurations that contained a spanning
cluster extending along the width of the strip $L$. Thus, $L$ sets
the appropriate length scale for the systems used, and the critical
interfaces can be studied by the theory of SLE in the scaling limit.
A turn right (or, alternatively, left) tie-breaking rule \cite{AS},
is used on the square lattice as a procedure to identify
the external perimeters (hulls) of the geometrical spin clusters
without any self-intersection. We note that in this case the hulls and
the external perimeters are the same.
To identify the interfaces in the upper half complex plane, a walker
moves along the edges of the dual lattice, starting from the origin
as sketched in figure \ref{figure2}. At the first step of the walk,
a spin (+) lies to the right of the walker (this direction is chosen
to be the preferable direction). After arriving at each site on the
dual lattice there are $3$ possibilities for the walker: it can
cross any of the $3$ nearest bonds of the original lattice. At the
first step of selection, it chooses the bonds containing two
different spins where crossing each of them leaves the spin $(+)$ to
the right and $(-)$ to the left of walker. The direction right or
left are defined locally according to the orientation of the walker.
If there are still two possibilities for crossing, the walker
chooses the bond which accords with the turn right tie-breaking
rule. It turns toward the bond which is on its right-bond side with
respect to its last direction in the last walk, if there is no
selected bond to its right, it prefers to move straight on and if
there is also no one there, it turns to its left. The procedure is
repeated iteratively until the walker touches the upper boundary.
The resulting interface is a unique one which has no
self-intersection and never gets trapped \cite{AS}. Note that we
just take the samples including a vertical spanning cluster in the
$y$-direction. The fractal dimension of the interfaces at
criticality, $D_{I}$, is obtained using the standard finite-size
scaling procedure. The length of interfaces is related to the sample
size as \be l\propto L^{D_{I}}.\label{eq3.1} \ee Indeed the fractal
dimension of the conformally invariant curves is provided by the SLE
theory as \be D_{I}= 1 + \frac{\kappa}{8} \label{eq3.2} \ee in which
diffusivity $\kappa$, as mentioned in the introduction,
characterizes different universality classes, and so does $D_I$. For
the regular Ising model, the diffusivity is believed to be $\kappa =
3$, and thus $D_{I}=\frac{11}{8}=1.375$. In addition, the fractal
dimension of the spanning spin cluster at $T_c$, obeys the relation
\be M\propto L^{D_{c}} \label{eq3.3} \ee where $M$ is the mass of
the cluster, and obtained by counting all the nearest-neighbor positive
(negative) spins to the right (left) of the SLE trace, as shown in figure \ref{figure2}. 
The exact value of $D_c$ for the regular Ising model is
$D_{c}=\frac{187}{96}=1.9479...$ \cite{DB}. Besides these, we find
the winding angle variance through the winding angle function $w(e)$
as defined by Wilson and Wieland \cite{WW}. For each edge on the
dual lattice, there is a value for the winding function $w(e)$ at
that edge such that the winding angle at the neighboring edge $e'$
is defined by $w(e') = w(e)$ + $the$ $turning$ $angle$ $from$ $e$
$to$ $e'$ $measured$ $in$ $radians$. It is shown that the variance
of the winding angles, grows with the sample size like  \be
\langle\theta^{2}\rangle= a + \frac{\kappa}{4} \ln(L).\label{eq3.4}
\ee Thus, by plotting $\langle\theta^{2}\rangle$ versus $\ln(L)$,
the slope gives a direct measure of $\kappa$ \cite{WW}.

The finite-size scaling of the spanning cluster size, is of the form
\cite{SA} \be M(L)=L^{D_{c}} \tilde{M}(L/\xi)\label{eq3.5} \ee where
the correlation length, also known as the connectedness length,
behaves like $\xi \sim (T-T_{c})^{-\nu_G}$, and the scaling function
$\tilde{M}(x)$ tends to a constant as its argument goes to zero at
$T_{c}$. Thus, the correlation length exponent $\nu_G$ of the
geometrical clusters, is estimated by a value that results in a data
collapse in a scaling plot $L^{-D_{c}}M(L)$ against
$L^{1/\nu_G}(\beta/\beta_c -1)$. Among other percolation quantities
of interest is the percolation strength $P_{\infty}(L)=M(L)/ L^2$,
which is the probability that a site chosen at random belongs to the
spanning cluster \cite{SA}. $P_{\infty}$ plays the role of an order
parameter for the percolation transition, and vanishes at the
percolation threshold $p_c=\tanh\beta_c$, at a rate specified by an
exponent $\beta_G$ defined by $P_{\infty}\propto
(p-p_c)^{\beta_G}$. The finite-size scaling relation for the
strength of percolation, is as follows \cite{SA}: \be P_{\infty}(L)
= L^{-\beta_G/\nu_G} \tilde{P}(L/\xi).\label{eq3.6}\ee Thus, at the
critical point $T_c$, a log-log plot of $P_{\infty}$ against $L$
must be a straight line with a slope equal to the ratio $-\beta_G
/\nu_G$. In the next section, we present our results for the
two-dimensional TBIM.

\section{Results and discussion}
\begin{figure}
\includegraphics{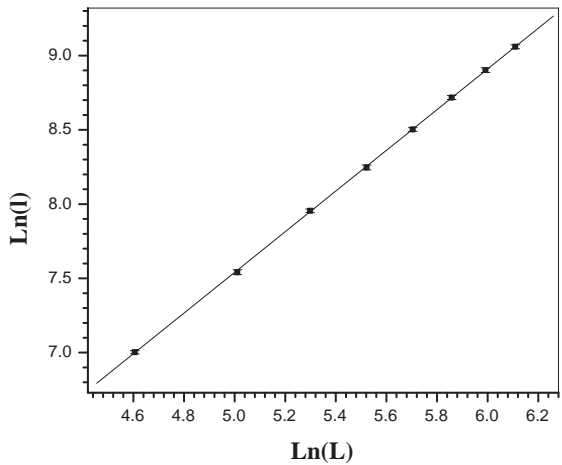}
\caption{A log-log plot of $l$ against $L$ for the two-dimensional
TBIM. The slope of the straight line gives the fractal dimension of
the external perimeters $D_{I}=1.373(8)$. The error bars are smaller
or comparable with the symbol size.} \label{figure3}
\end{figure}

In this section we present and discuss our main results for the
two-dimensional TBIM as obtained from simulations based on the
methods pointed out in the previous section. We performed
simulations for eight different system sizes $L=$ 100, 150, 200,
250, 300, 350, 400, and 450. Only the spin configurations including
a vertical spanning cluster are considered for analysis, and the
statistical errors were estimated by means of binning the
accumulated data. As it appears in figure \ref{figure3}, the slope
of a log-log plot of the spanning length $l$ versus the system size
$L$, results in a fractal dimension $D_{I}=1.373(8)$, which is
equivalent to a $\kappa =2.984(17)$ as given by equation
(\ref{eq3.2}).

\begin{figure}
\includegraphics{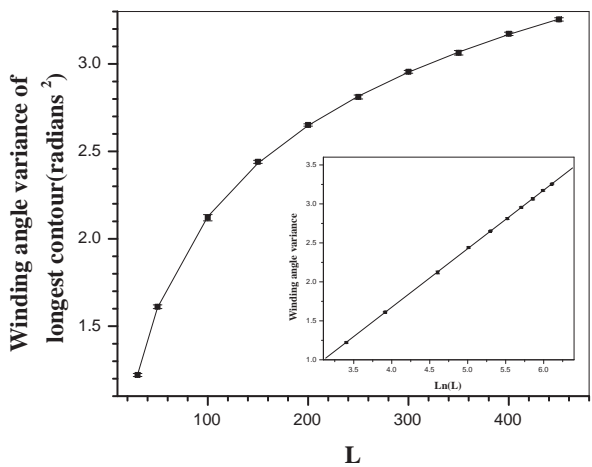}
\caption{A plot of $\langle\theta^{2}\rangle$ versus $L$ for the
two-dimensional TBIM. In the inset, the variance is in
semi-logarithmic coordinates. The error bars are smaller or
comparable with the symbol size.}\label{figure4}
\end{figure}

To confirm our results, we also measured the winding angle variance
along the spanning contour by performing simulations for 10
different system sizes $L=$ 30, 50, 100, 150, 200, 250, 300, 350,
400, and 450. A plot of $\langle\theta^{2}\rangle$ against $L$ is
shown in figure \ref{figure4}. A curve of the form $a_{1}+ a_{2}
\ln(L)$ with parameters $a_{1}=-1.330(10)$ and $a_{2}=0.751(2)$
(where $a_{2}=\frac{\kappa}{4}$), was least-squares fitted to these
data. Furthermore, plotting $\langle\theta^{2}\rangle$ versus
$\ln(L)$, results in a SLE parameter $\kappa =3.004(9)$ as
illustrated in the inset of figure \ref{figure4}.


\begin{figure}
\includegraphics{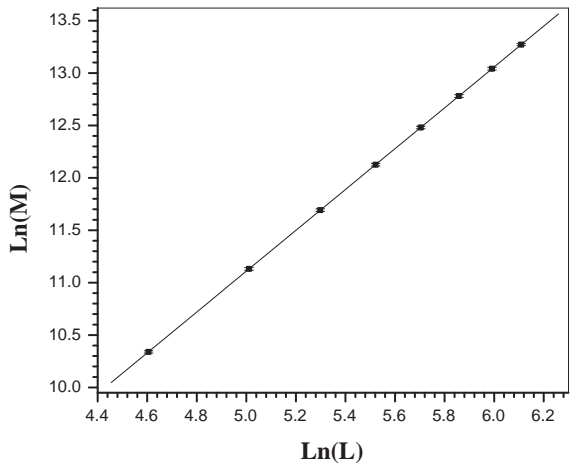}
\caption{A log-log plot of $M$ against $L$. The slope of the
straight line gives the fractal dimension of the geometrical
clusters at $T_c$, $D_{c}=1.948(3)$.} \label{figure5}
\end{figure}

A log-log plot of the spanning cluster mass $M$ versus the system
size $L$ is shown in figure \ref{figure5}. The slope gives a fractal
dimension $D_{c}=1.948(3)$.

\begin{figure}
\includegraphics{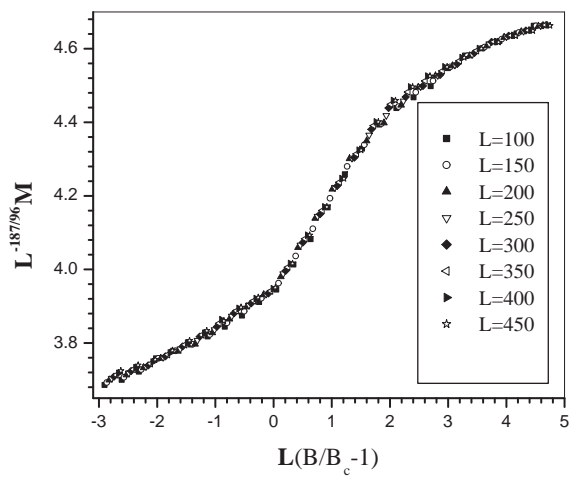}
\caption{A scaling plot for the spanning cluster mass.  By varying
$\nu_G$, and evaluating the quality of the data collapse, our best
estimate of the correlation length exponent is $\nu_G = 1.01(2)$}
\label{figure6}
\end{figure}

\begin{figure}
\includegraphics{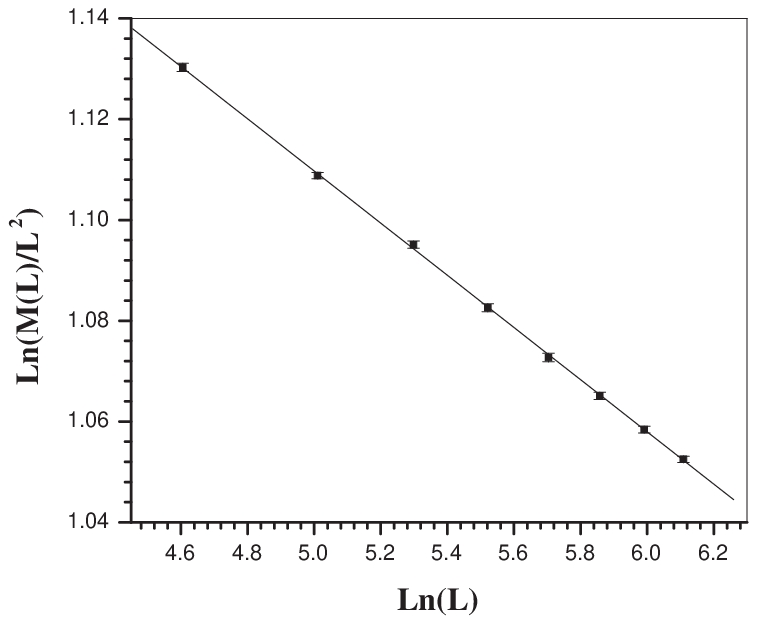}
\caption{A log-log plot of $P_{\infty}$ against $L$. The slope of
the straight line gives $\beta_G = 0.051(3)$} \label{figure7}
\end{figure}

Using the finite-size scaling ansatz as given in equation
(\ref{eq3.5}), the correlation length exponent $\nu_G$ for the
emerging spanning cluster, is estimated from a scaling plot of $
L^{-D_c}M(L)$ against $L^{1/\nu_G}(\beta/\beta_c-1)$ as shown in
figure \ref{figure6}. By varying $\nu_G$, and evaluating the quality
of the data collapse, our best estimate of the correlation length
exponent for the geometrical clusters is $\nu_G = 1.01(2)$. Finally,
the slope of the log-log plot of $P_{\infty}$ versus $L$ results in
$\beta_G=0.051(3)$, as shown in figure \ref{figure7}, which fits
well into the hyperscaling relation for the percolation exponents in
$d$ spatial dimensions $D_{c}=d-\frac{\beta_G}{\nu_G}$.

The obtained results for the two-dimensional TBIM are listed in
table \ref{table1} for comparison with the analytical results of the
regular Ising model. At criticality, the fractal dimension of the
spin clusters $D_{c}$ and interfaces $D_{I}$  have been found to be
consistent with the analytical results obtained for the regular
Ising model, despite the temperature-dependent annealed bond dilution.

\begin{table}
\caption{\label{table1}The critical exponents of the geometrical clusters in $2d$ TBIM are
compared with those obtained for $2d$ regular Ising model.}
\begin{indented}
\item[]\begin{tabular}{llllll}
\br

                      &$\nu_G$ &$\beta_G$ & $\kappa$ &  $ D_{I}$  &$D_{c}$ \\ \hline
TBIM &1.01(2)& 0.051(3) & 3.004(9) & 1.373(8) & 1.948(3) \\
Ising Model & 1.00 & $5/96$ & 3& 1.375 & $187/96$ \\
\end{tabular}
\end{indented}
\end{table}


It must be noted that although the geometrical clusters (i.e., the
neighboring sites of the same spin sign) uniquely characterize the
universality class of the critical system, they do not, however, represent the
critical droplets. Hence, the percolation critical exponents of the
geometrical clusters, do not in general coincide with those of the
corresponding thermal quantities. The critical droplets, are more
precisely specified by the so-called Fortuin-Kastelyn (FK) clusters whose
diffusivity parameter $\kappa$ has a duality relation with that of
the geometric spin clusters \cite{CK80,FK,DB1}. The FK clusters can
be obtained from the geometrical clusters through a random
decimation of bonds by a suitable probability, in this case
$1-\tanh(\beta)$, and are therefore less compact. Thus, despite the
correlation length exponent $\nu_G =1$, we note that the geometrical
clusters are too compact to represent the critical droplets, and the
exponent $\beta_G = 5/96$ differs appreciably from the corresponding
thermal exponent $\beta =1/8$, associated with the vanishing of the
magnetization order parameter at $T_c$. The value of the correlation
length (also known as the connectivity length) exponent $\nu_G= 1$,
for the geometrical clusters of the two-dimensional TBIM, is in
excellent agreement with the values obtained from a real-space
renormalization group analysis \cite{CK80,Co}, high-temperature
series expansion studies \cite{SG}, and precision numerical
simulations of the geometrical clusters of the standard
two-dimensional Ising model \cite{For,JS}. This result, however, opens
a question about a near perfect collapse onto a universal
function for the same data obtained for the regular Ising model, but
with a different exponent $\frac{15}{8}$ \cite{AS}.

As can be seen from table \ref{table1}, within the statistical
uncertainty, the value of the SLE parameter $\kappa$, the fractal
dimensions, and the percolation exponents of the geometrical spin
clusters of the two-dimensional TBIM, are in excellent agreement
with those of the corresponding regular Ising model. These results
agree well with an earlier study of the thermodynamic critical
behavior of the two-dimensional TBIM, which places the model system
in the universality class of the standard two-dimensional Ising
model \cite{DM}, and the Fisher renormalization relations, which
assert that annealed bond dilution can only change the critical
exponents if the specific heat exponent $\alpha_r$ of the regular
model is positive ($\alpha_r > 0$) \cite{Fis}. The
two-dimensional regular Ising model, however, is characterized by a
logarithmic divergence of the specific heat, $\alpha_r=0$, and
the exponents remain unchanged.

As for the fractal behavior of the geometrical clusters away from
the criticality, we note that for all temperatures below the
transition point $T<T_c$ (or $p>p_c$),  $D_{\rm c}$ must equal the
space dimensions $d=2$, otherwise the percolation strength would
vanish in the thermodynamic limit of an infinite lattice. At $T_c$
($p=p_c$), the geometrical clusters of the two-dimensional TBIM, are
fractal with a fractal dimension $D_c = 1.948(3)$ ($<2$), thus
rendering the percolation strength zero at the critical point, as
expected. Hence, one expects the fractal dimension of the
geometrical clusters to change discontinuously from $D_c=d=2$ for
$T<T_c$, to $D_c=1.948(3)$ at $T_c$. This expectation is validated
by the data of reference \cite{AS}, where the so-called `effective'
fractal dimensions undergo a sharp crossover at $T_c$. We believe
that the crossover is a finite-size effect, and a remnant of the
discontinuity at $T_c$ in the thermodynamic limit. As for the
temperatures above the critical point $T>T_c$, there are no spanning
geometrical clusters in the thermodynamic limit and the procedures
used here become inapplicable. However, other standard procedures
such as the box counting method may be employed to investigate the
fractal behavior of the geometrical clusters above $T_c$ and within
a region of linear size of the order of the finite correlation
length $\xi$.

\section{Summary}
The fractal behavior of the geometrical spin clusters, are obtained
for the thermalized bond  Ising model in two dimensions. For this
purpose, a modified Wolff single-cluster Monte Carlo simulation is
used to generate equilibrium spin configurations on square lattices
in the critical region. The obtained values for the fractal
dimensions of the spanning geometrical clusters, $D_{c}$, and that
of their interfaces, $D_{I}$, are in perfect agreement with those
reported for the regular Ising model. The variance of the winding
angles results in a value $\kappa=3.004(9)$ for the SLE parameter,
thus placing it in the universality class of the regular Ising
model. Furthermore, the percolation exponents of the geometrical
spin clusters at $T_c$, are  found to be consistent with those
reported for the regular Ising model. These consistencies are
explained in terms of the Fisher renormalization relations, which
express the thermodynamic critical exponents of systems with
annealed bond dilution in terms of those of the regular model
system.

\end{document}